\documentclass[journal]{IEEEtran}
\usepackage{cite}
\usepackage{amsmath,amssymb,amsfonts}
\usepackage{algorithmic}
\usepackage{graphicx}
\usepackage{textcomp}
\usepackage{multirow}
\usepackage{booktabs}
\usepackage{amsthm}
\newtheorem{definition}{Definition}
\usepackage{array}
\usepackage{makecell}

\def\BibTeX{{\rm B\kern-.05em{\sc i\kern-.025em b}\kern-.08em
    T\kern-.1667em\lower.7ex\hbox{E}\kern-.125emX}}
\begin{document}

\title{Typical Scenarios Generation Method Considering System-level Characteristics of Power System}
\author{Tao Li, \IEEEmembership{Graduate Student Member, IEEE}, Chen Shen, \IEEEmembership{Senior Member, IEEE}
\thanks{This work was supported by Joint Fund Project supported by the National Natural Science Foundation of China (U2166601) (Corresponding author: Chen Shen.)}
\thanks{The authors are with the Department of Electrical Engineering, Tsinghua
University, Beijing 100084, China (e-mail: t-li24@mails.tsinghua.edu.cn;
shenchen@mail.tsinghua.edu.cn}}
\maketitle

\begin{abstract}
This paper proposes a method for generating typical scenarios based on system-level macroscopic characteristics of power system and considering its stability properties. First, considering uncertainties such as renewable energy generation in power-electronics-dominated power systems, multidimensional scaling is used to construct an electrical coordinate system. Based on this, system-level characteristics of the distribution of physical quantities, such as power generation and load, are characterized. Furthermore, a method for generating typical scenarios based on the system's system-level characteristics and stability properties is proposed. For the obtained joint probability distribution of system-level characteristics, weighted Mahalanobis distance can be used to predict the stability properties of random scenarios. Finally, the typicality and representativeness of the scenarios generated by the proposed method with respect to stability properties are verified on the CSEE benchmark case, and stability prediction for random scenarios is achieved using a probabilistic testing method.
\end{abstract}

\begin{IEEEkeywords}
Gaussian mixture module, multi-dimensional scaling, system-level characteristics, typical scenarios generation.
\end{IEEEkeywords}

\section{Introduction}
\label{sec:introduction}
\IEEEPARstart{W}{ith} the large-scale grid connection of renewable energy sources such as wind power and photovoltaics, power systems face increasingly severe uncertainty challenges. Traditional power systems often conduct safety and stability analyses for some representative and deterministic operating modes during planning and operation, but cannot fully consider the changes in operating status caused by fluctuations in renewable energy generation and the potential risks that may arise after such changes \cite{a1}. Therefore, constructing typical scenarios to characterize the system's operating characteristics under different output conditions has become a key component of operation and control of power-electronics-dominated power systems. 

State Grid Corporation of China uses "operating mode" to characterize power systems' operating characteristics under generation-load balance and across various load conditions. Correspondingly, in power system research, "scenario" refers to a specific combination of values for key external variables, such as power generation, load demand, and meteorological conditions, at a given time scale \cite{a2}. In summary, the two widely used concepts, operating mode and scenario, share similar connotations and are both sampling points of a series of random variables. However, in practice, evaluation of operating modes not only considers power flow convergence characteristics and distribution rationality, but also requires comprehensive consideration of dynamic characteristics, including the abundance of adjustable resources \cite{a3,a4}. Power system safety and stability guidelines \cite{a5} and Shandong province power system dispatch management \cite{a6}, among other power system operation guidance documents, also specify detailed requirements for the dynamic characteristics of operating modes, including the ability to withstand disturbances under N-1 fault conditions.

In summary, traditional concepts and analysis methods for operating modes and scenarios no longer fully meet the dynamic requirements of real power systems. It is necessary to consider the dynamic response characteristics of power electronic components, such as renewable energy sources, based on power flow analysis, and to broaden the connotation of "scenario". Based on the existing definitions of operating modes and scenarios, and considering the requirements for dynamic characteristics in actual power system operation, we define "scenario" as:
\begin{definition}
A scenario is a representative operating condition of the power system that encapsulates its steady-state power balance, power flow distribution, and associated dynamic behavior under disturbances.
\end{definition}

Currently, scenario generation methods can be divided into three types: sampling-based, prediction-based, and optimization-based \cite{a7}. Sampling-based methods generate a large number of discrete scenarios by sampling from the probability distribution \cite{a7.1,a7.2}, but they are highly dependent on the accuracy of the distributional model. Prediction-based methods generate continuous scenarios across multiple time scales by learning patterns of historical evolution that characterize temporal correlations, but they are highly sensitive to data quality and model training, and they lack explicit probabilistic interpretations. Optimization-based methods select a small number of typical scenarios from a large set of candidate samples by optimizing distance indices, such as moment matching, clustering, backward reduction, and forward selection \cite{a8}, which achieve high computational accuracy but suffer from low computational efficiency and limited scalability.

Existing typical scenario generation methods are essentially optimization-based, which generate scenarios via random sampling or measurement, then select the most "typical" scenarios from the set. Different methods imply different definitions of "typical": 1) "typical" defined based on background knowledge, which involves classifying scenarios according to holidays or quarters, and then making subjective selections \cite{a6}; 2) "typical" definition based on minimizing probability distance, which involves iteratively minimizing probability distance to select the optimal combination of a given number of typical scenarios \cite{a9,a11}; 3) "typical" definition based on clustering information, which is essentially a distance-based optimization method. However, unlike iterative methods, clustering information implicitly incorporates scenario classification and intra-class scenario averaging \cite{a12,a13}. 

Scenarios implied by the above methods include only sets of generation and load values at the component level, defined as the \emph{"microscopic level"}; whereas the \emph{"macroscopic level"} discussed in this paper corresponds to the distributional mapping of this set in a given embedding space. The introduction of the "macroscopic level" (i.e., distribution characteristics of physical quantities) is motivated by its strong correlation with the essence of dynamic and power flow characteristics, such as local source–load consumption and low-inertia conditions, which are closely related to system dynamics and stability margins. To analyze power systems and source–load distribution characteristics at the macroscopic level, it is necessary to establish a mathematical structural model of power systems.

A widely used power system structure modeling method is Graph Neural Networks(GNN)\cite{a13.1}. GNNs can automatically extract high-dimensional features from graph-structured data and are widely used in power system transient stability assessment and dynamic characteristic prediction \cite{a14,a15,a16}. Whereas the exploration of system structure, feature modeling, and scenario analysis remains relatively limited. Another concept that reflects the structural features of power systems is electrical distance, which measures the electrical coupling between nodes \cite{a17} and is usually based on impedance and similarity calculations, reflecting only "distance" without a clear spatial structure. However, electrical distance alone cannot intuitively capture the spatial relationships among nodes, which limits the applicability of cluster analysis and fault visualization. Therefore, mapping electrical distance onto geometric coordinates while preserving electrical proximity yields an electrical coordinate system that is a spatial extension of electrical distance information \cite{a18}. Reference \cite{a19} obtains a two-dimensional pseudo-geographical representation of the power system using the multi-dimensional scaling (MDS) algorithm, but focuses on its comparison with geographical coordinates; References \cite{a20,a21} use a greedy algorithm to determine the reference node based on electrical distance and construct an electrical coordinate system by distance of the reference node, but the complexity of high-dimensional calculation is high.

To address the above problems, this paper proposes a method for constructing typical scenarios of power grids by integrating system-level macroscopic characteristics and system dynamics. The specific innovations are as follows:
\begin{enumerate}
\item A method for constructing an electrical coordinate system based on MDS is proposed, achieving high-precision characterization of electrical distances while preserving some Euclidean geometric characteristics of the power grid topology.

\item Based on the electrical coordinate system, system-level macroscopic characteristics reflecting the distribution characteristics of quantities such as generation and load are defined, and their correlation with system stability is verified.

\item A method for generating typical scenarios based on system-level characteristics is proposed, which generates scenarios that reflect the stability characteristics of power-electronics-dominated power systems, and stability prediction for random scenarios is achieved using probability testing.
\end{enumerate}

The structure of this paper is as follows: Section \ref{111} introduces a method for constructing an electrical coordinate system based on MDS and analyzes its applicability in power systems; Section \ref{222} defines macroscopic characteristics reflecting system-level features of power system based on the electrical coordinate system and verifies their correlation with system transient stability; Section \ref{333} proposes a method for constructing typical scenarios of hybrid power grids based on system-level characteristics of the system and gives its application method in predicting system stability; Section \ref{444} verifies the effectiveness of the proposed method in the CSEE standard examples published by the Chinese Society for Electrical Engineering.
\section{Construction of the Electrical Coordinate System}\label{111}
\subsection{Definition of Electrical Distance}
There are two main methods to measure node connectivity in power systems: one is to use indices such as sensitivity and transfer distribution factor to measure the mutual influence between two components; the other is to use the distance matrix, where electrical distance between nodes is only related to the physical connection between components and is not affected by power, load, or other operating conditions. Reference \cite{a22} first defined the electrical distance based on the impedance relationship between nodes, whereas for large-scale power systems it can be defined using the node impedance matrix Z, as shown in \eqref{eq1}.
\begin{equation}D_{a,b}=|Z_{a,a}+Z_{b,b}-Z_{b,a}-Z_{a,b}| \label{eq1}\end{equation}
Where: $D$ represents the distance matrix of system, and $D_{a,b}$ represents the electrical distance between node $a$ and node $b$.

Mathematically, distance definition has three requirements:
\begin{enumerate}
    \item Non-negativity and identity: $d(x,y)\geq0$, and $d(x,y)=0$ if and only if $x=y$;
    \item Symmetry: $d(x,y)=d(y,x)$;
    \item Triangle inequality: $d(x,y)\leq d(x,z)+d(z,y)$.
\end{enumerate}

It is easily verified that the definition of \eqref{eq1} satisfies the first two of the above three requirements, while the complex characteristics of power systems lead to non-Euclidean characteristics of electrical distance, which barely satisfy the triangle inequality.
\subsection{MDS Principles and Applicability in Power Systems}
Electrical distance can only reflect the relative distance between two buses and cannot determine the absolute position of each bus, so it fails to capture the system's overall characteristics. The coupling relationship between nodes exhibits high-dimensional complexity in both mathematical properties and network structure, and must be reduced to obtain a low-dimensional representation suitable for constructing system-level characteristics. Therefore, the MDS algorithm is used to construct an electrical distance coordinate system \cite{a19}, thereby reducing the dimension of the high-dimensional space of the power system. 

The classic MDS algorithm is based on Schoenberg's theorem, applied to Euclidean space, which converts the distance matrix into an inner product matrix and then obtains low-dimensional coordinates via eigenvalue decomposition. Specifically, $D$ is the distance matrix of the system, and the inner product matrix B can be obtained through the double-centering transformation \cite{a23}: 

\begin{equation}B=-\dfrac{1}{2}CD^{(2)}C,C=I-\textbf{1}\textbf{1}^T \label{eq2}\end{equation}
Where: $D^{(2)}$ is the distance matrix after element-wise squaring, and $C$ is the centering matrix. Matrix $B$ is the Gram matrix in Euclidean space. The first $k$ positive eigenvalues and corresponding eigenvectors of matrix $B$ are taken to construct the coordinate vector, that is, the coordinates of $D$ in k-dimensional Euclidean space, as shown in \eqref{eq3}. 
\begin{equation}B=V\Lambda V^T , X_k=V_k\Lambda_k^{1/2}\label{eq3}\end{equation}
Where: $\Lambda_k$ is a diagonal matrix composed of the first $k$ positive eigenvalues, and $V_k$ is the corresponding eigenvector.

Therefore, classical MDS is strongly correlated with positive eigenvalues of the matrix  $B$. The number of positive eigenvalues determines the dimension of the maximal mappable space, and the magnitudes of the eigenvalues determine the influence of that dimension on the overall coordinate system. The non-Euclidean nature of electrical distance yields negative eigenvalues, thereby directly reducing the accuracy of the mapping. Therefore, for highly non-Euclidean systems with a large number of negative eigenvalues, the accuracy of classical MDS has a theoretical upper limit.

Non-classical MDS includes metric MDS and non-metric MDS, which are essentially optimization algorithms that minimize the Stress function \cite{a23}. Compared with non-metric MDS, metric MDS pays more attention to the restoration of specific values, and the stress function is shown in \eqref{eq4}. Existing optimization methods generally find the global minimum of functions via iterative methods, such as the SMACOF (Scaling by Majorizing a Complicated Function) algorithm \cite{a24}.
\begin{equation}\sigma_r=\sum_{i<j}w_{ij}(D_{ij}-d_{ij})^2 \label{eq4}\end{equation}
Where: $D_{ij}$ represents the input distance matrix elements; $d_{ij}$ is the Euclidean distance between nodes in a low-dimensional space. $w_{ij}$ is the weight, which generally can be set to 1.

The accuracy of metric MDS in low-dimensional spaces is not limited by the non-Euclidean nature of electrical distances, thereby preserving the specific values of distances.

\subsection{Characteristics of the Electrical Coordinate System}

The fundamental purpose of establishing an electrical coordinate system is to precisely depict the overall distribution of nodes in the entire system in a low-dimensional space based on electrical distances. Therefore, by defining a new Euclidean distance matrix $d$ using the coordinate system and calculating the Pearson correlation between $d$ and the actual electrical distance matrix $D$, the accuracy of the electrical coordinate system can be quantitatively analyzed, as shown in \eqref{eq5}.
\begin{equation}\rho=\dfrac{cov(d,D)}{\sigma_d\sigma_D}\label{eq5}\end{equation}

\begin{figure}[!tbp]
\centerline{\includegraphics[width=\columnwidth]{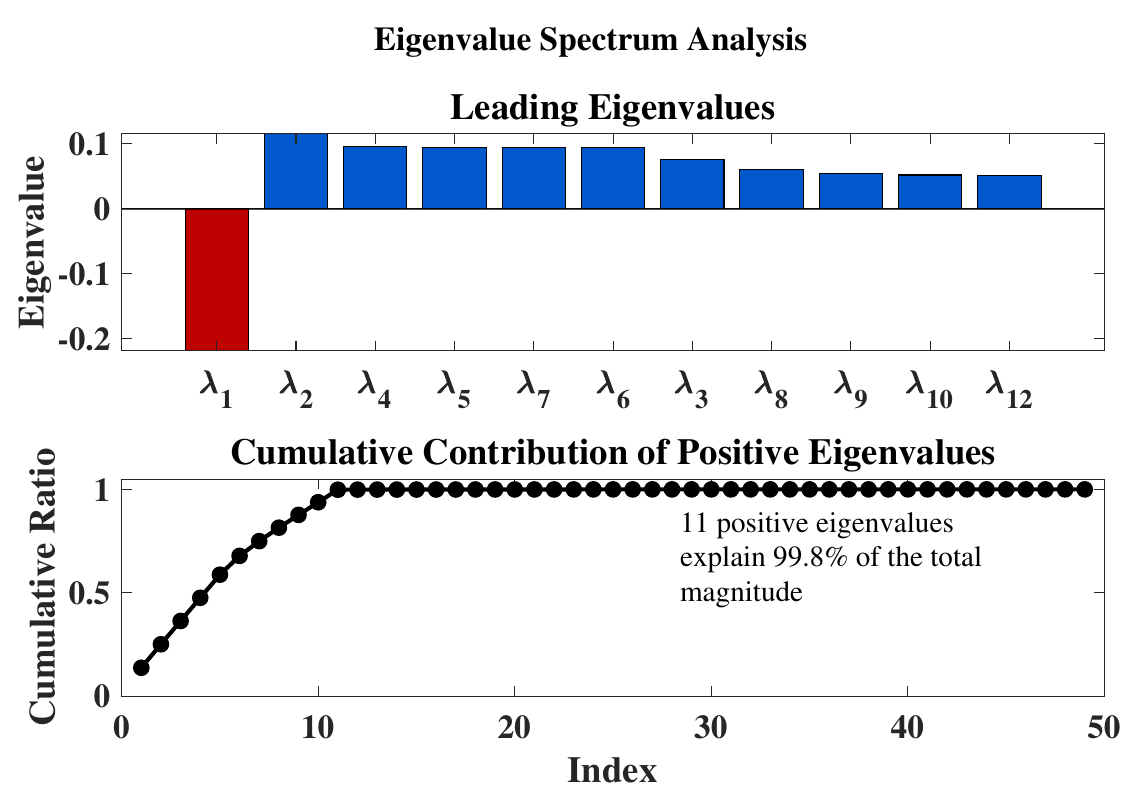}}
\caption{Eigenvalue spectrum of the benchmark case}
\label{fig1}
\end{figure}

Tests were conducted in CSEE benchmark case\cite{a25}, which has 93 nodes. After calculation, there are 49 groups of positive eigenvalues, and the specific numerical distribution is shown in Figure \ref{fig1}. The first 11 groups of feature roots account for 99.8\% of the proportion. Therefore, according to the characteristics of the MDS algorithm, the first 11 dimensions are dominant. In contrast, non-Euclidean spaces with more than 49 dimensions will distort the electrical coordinate system. Therefore, theoretically, the optimal dimensionality is 11.

\begin{figure}[tbp]
\centerline{\includegraphics[width=\columnwidth]{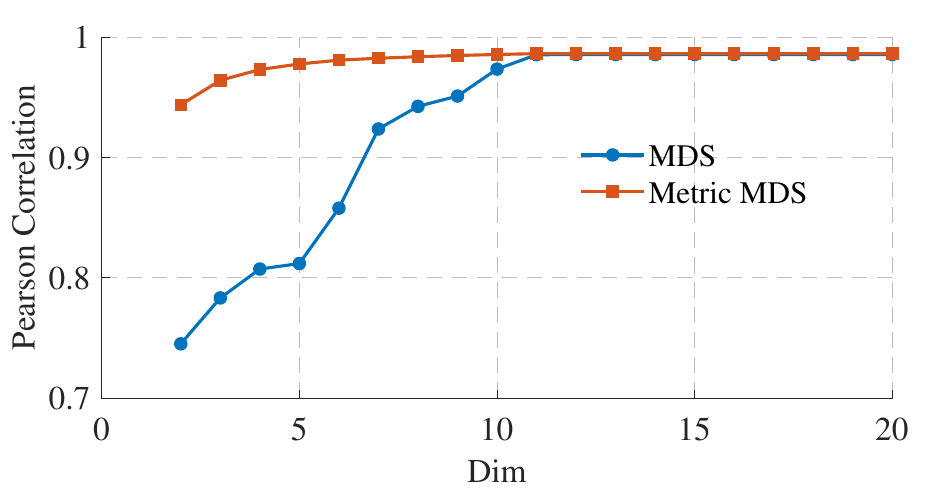}}
\caption{Pearson correlation–spatial dimension relationship}
\label{fig2}
\end{figure}

Set the mapping dimensions from 2 to 20, and the Pearson correlation between the calculated electrical coordinate system distance matrix and the original electrical distance matrix is shown in Figure \ref{fig2}. As the dimension increases, the correlation improves, and the accuracy of measuring MDS is significantly higher than that of classical MDS. When the dimension exceeds 11 dimensions, the correlation coefficients of both the classical MDS and the metric MDS converge to 0.985.


The correlation of the classical MDS in two-dimensional space exceeds 0.75 (a correlation above 0.70 is generally considered strong), indicating that the low-dimensional space at this time can already accurately describe the coupling relationships among nodes in the power system. Meanwhile, since the accuracy of classical MDS is limited by eigenvalues, the mapping accuracy of the metric MDS in the low-dimensional space is significantly higher than that of classical MDS.

\section{Definition and analysis of system-level characteristics of the system}\label{222}
Based on the electrical coordinate system defined in Section \ref{111}, the distribution characteristics of variables such as power generation, system inertia, and load are considered to define the system-level characteristics. These characteristics directly affect the rationality of the power flow distribution and the stability margin of the power system. For instance, an uneven power-load distribution may result in excessive power flow in some transmission lines, thereby reducing the system stability margin. Therefore, there is a correlation among system-level characteristics, power flow, and stability, which is an important reference for evaluating the system's operational safety.

\subsection{Definition of System-level Macroscopic Characteristics}
Based on engineering experience, system stability is strongly correlated with parameters such as generation-load distribution and inertia distribution. Based on the electrical coordinate system, the statistical characteristics of various physical quantities, including generation and load, are quantified. The specifically defined indices are shown in Table \ref{tab1}. In Table\ref{tab1}, norm-based matching refers to numerical difference in the distribution of two physical quantities, which is obtained by the maximum element of the matrix, as shown in \eqref{eq6}; Numerical difference can better depict the influence of extreme data, but are greatly affected by computational details such as dimension and normalization, and cannot thus describe the system-level characteristics. Structural similarity is defined using the Structural Similarity Index Measure (SSIM) to describe the similarity of the distribution characteristics of two physical quantities at the global level, as shown in \eqref{eq7}.
\begin{equation}||I_1-I_2||_{\infty}=max(abs(I_1-I_2))\label{eq6}\end{equation}
\begin{equation}SSIM(I_1,I_2)=\dfrac{2\mu_1\mu_2+C_1)(2\sigma_{1,2}+C_2)}{(\mu_1^2+\mu_2^2+C_1)(\sigma_1^2+\sigma_2^2+C_2)}\label{eq7}\end{equation}

\begin{table}
\caption{Definition of System-Level Characteristics}
\setlength{\tabcolsep}{3pt}
\begin{tabular}{>{\centering\arraybackslash}p{12pt}>{\centering\arraybackslash}p{70pt}>{\centering\arraybackslash}p{95pt}>{\centering\arraybackslash}p{50pt}}
\hline
Num &
Name &
Physical meaning& Symbol\\
\hline
1 & active generation-load norm-based matching and SSIM & It affects the heterogeneous distribution of power flow and thereby influences stability problems such as fault propagation\cite{a26,a27}.& \makecell{$P_G-P_L$ \\norm \& SSIM} \\ \addlinespace
2 & reactive generation-load norm-based matching and SSIM & It affects the transient voltage response and simultaneously influences power flow convergence \cite{a28}.&\makecell{ $Q_G-Q_L$\\norm \& SSIM} \\ \addlinespace
3 & power electronic devices-reactive power generation SSIM & Measure whether the reactive power required by power electronic converters is consistent with generation. & \makecell{$P_{re}-Q_G $\\SSIM}\\ \addlinespace
4 & inertia-active load SSIM & \multirow{2}{115pt}{Measure the inertial support of the system.}& \makecell{$J-P_L$\\ SSIM}\\ 
5 & synchronous-renewable resources norm-based matching and SSIM &  & \makecell{$P_{sync}-P_{re} $\\norm \&SSIM}\\ \hline
\end{tabular}
\label{tab1}
\end{table}

\subsection{Correlation Analysis between System-level Characteristics and Stability}
\begin{figure}[!t]
    \centering
        \begin{minipage}{0.5\textwidth}
    \centering
    \includegraphics[width=\linewidth]{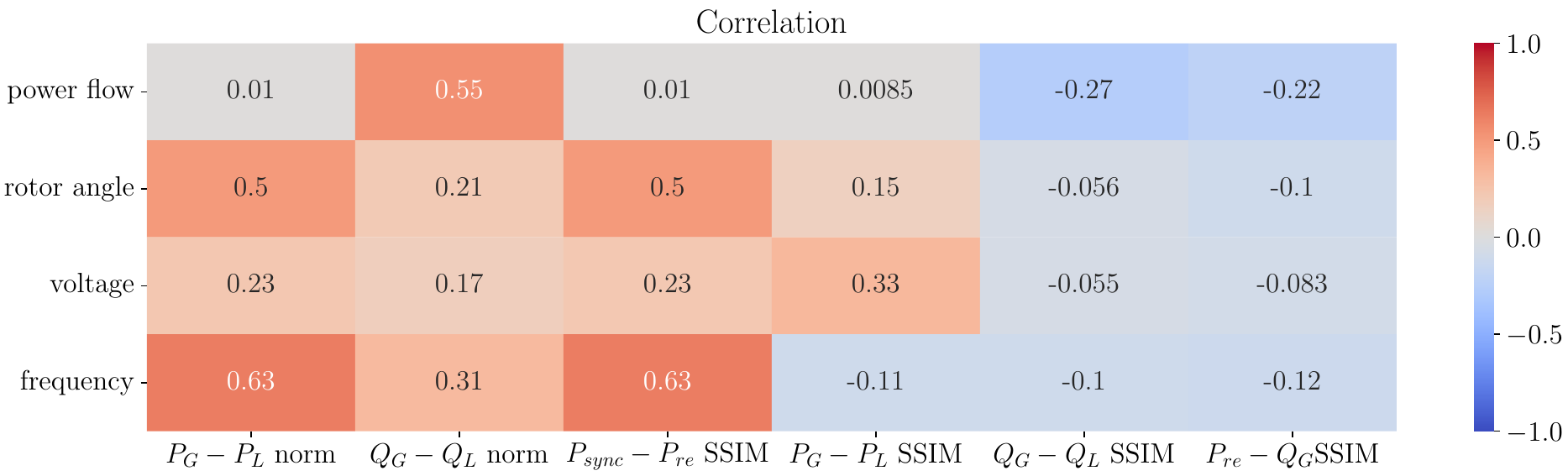}
    \\ (a) Pearson correlation
    \end{minipage}
    \hfill
    \begin{minipage}{0.5\textwidth}
    \centering
    \includegraphics[width=\linewidth]{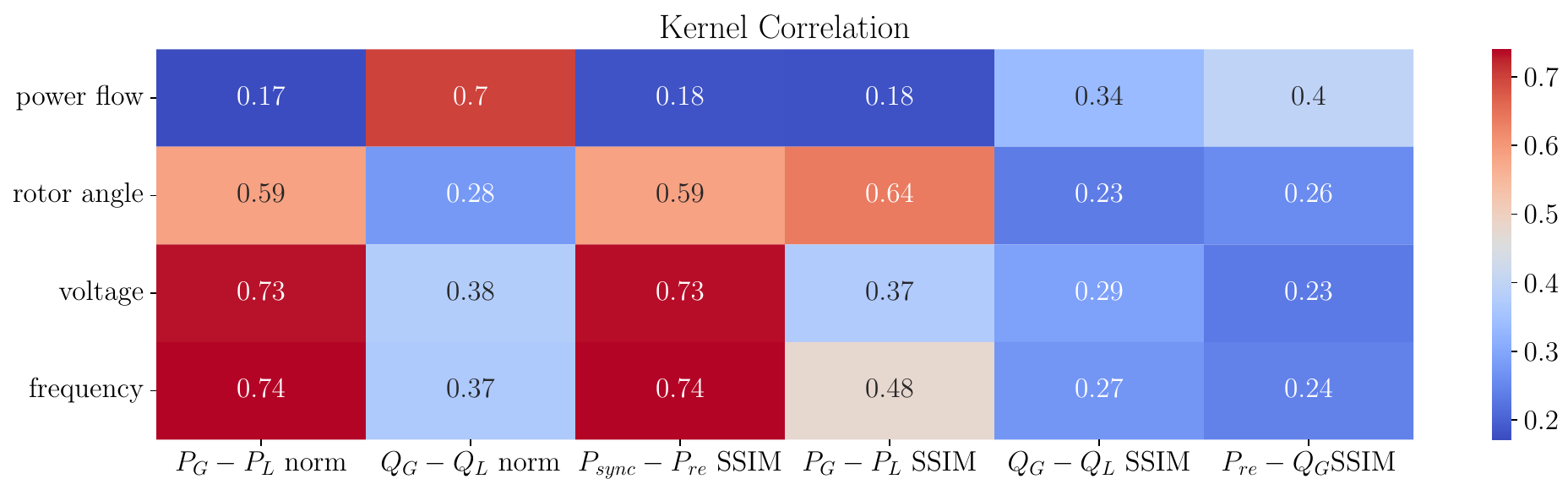}
    \\ (b) Kernel correlation
    \end{minipage}
        \vspace{0.8em}
        \begin{minipage}{0.5\textwidth}
    \centering
    \includegraphics[width=\linewidth]{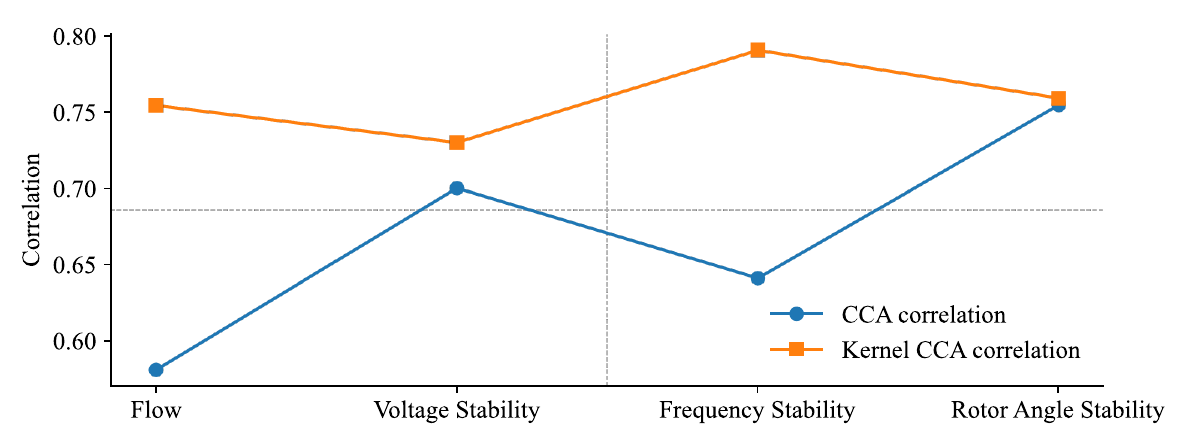}
    \\ (c) CCA and KCCA correlation
    \end{minipage}
    \hfill
    \caption{Correlation between system-level characteristics and system stability}
    \label{fig3}
\end{figure}

The system-level macroscopic characteristics are related to stability and thus affect the transient response under fault conditions. Calculating Pearson correlations between the defined system-level characteristics and the electrical quantities across different stability types quantifies the extent to which these characteristics affect system stability. Results are shown in Figure \ref{fig3}(a), where norm-based matching is positively correlated with stability indices, indicating that when the generation-load distribution does not match, stability problems are more likely to occur. Among them, the matching of synchronous generation with inverter-based resources and the distribution of active power generation and loads have a greater impact on frequency and rotor angle stability.

However, linear correlation cannot fully capture the relationship between system-level characteristics and system stability. The kernel function $K$ maps the original data into a high-dimensional space, capturing nonlinear coupling among variables. The correlation matrix between system-level characteristics and stability indices, computed using kernel functions, is shown in Figure \ref{fig3}(b). Compared with linear correlation, kernel functions more effectively capture nonlinear correlations between variables. Among them, there is a strong correlation between the active power generation-load distribution and rotor angle and frequency stability, and between the reactive power generation-load distribution and voltage stability.

To assess the combined influence of multiple system-level characteristics on stability, Canonical Correlation Analysis (CCA) is used to identify the maximum correlation between linear combinations of characteristics and stability indices. Let $X$ represent system-level characteristics, and $Y$ represent stability indices. CCA optimizes the coefficients $a$ and $b$ to maximize the correlation, as shown in \eqref{eq8}.
\begin{equation}\text{arg}\max_{a,b}\dfrac{cov(Xa,Yb)}{\sqrt{var(Xa)}\sqrt{var(Yb)}}\label{eq8}\end{equation}

Kernel CCA (KCCA) introduces a kernel function $K$ that maps the original variable sets $X$ and $Y$ into a high-dimensional feature space via a nonlinear mapping, and then performs CCA in this space. KCCA takes $\alpha$ and $\beta$ as the optimization parameters, as shown in \eqref{eq9}.
\begin{equation}\text{arg}\max_{\alpha,\beta}\dfrac{\alpha^TK_XK_Y\beta}{\sqrt{\alpha^T(K_X^2+\kappa I)\alpha}\sqrt{\beta^T(K_Y^2+\kappa I)\beta}}\label{eq9}\end{equation}
Among them, $K_X$ and $K_Y$ are the kernel matrices of the input data, and $\kappa$ is the regularization parameter. KCCA can capture nonlinear dynamical behaviors of the system, such as fluctuations of renewable energy and the impact of power-electronic control on system stability. Comprehensive correlations between stability indices and all system-level characteristics are calculated using linear CCA and kernel CCA, as shown in Figure \ref{fig3}(c). It can be observed that system-level characteristics exhibit strong correlations (over 0.7) with most stability indices and power flow convergence, and a few show relatively strong correlations (between 0.6 and 0.7), indicating that the defined system-level characteristics can reflect aspects of power system stability.

\section{typical scenarios generation method based on system-level characteristics}\label{333}
\begin{figure*}[tbp]
\centerline{\includegraphics[width=1.8\columnwidth]{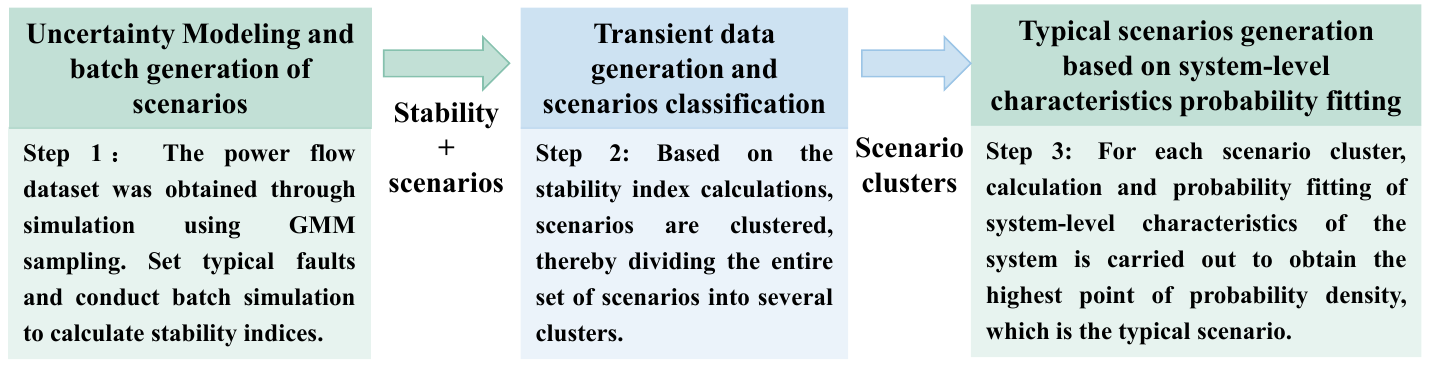}}
\caption{Flowchart of typical scenario generation}
\label{fig4}
\end{figure*}

Based on the system-level characteristics defined in Section \ref{222}, a new method for generating power system typical scenarios is proposed in this section. Classify power flow scenarios by their response characteristics under typical fault conditions. For each scenario cluster, the typical scenario is obtained by fitting system-level characteristics to Gaussian Mixture Model (GMM). The process is shown in Figure \ref{fig4}. The entire typical scenario generation process can be divided into three steps, as shown in the figure: Uncertainty Modeling and batch generation of scenarios, Transient data generation and scenarios classification, and Typical scenarios generation based on system-level characteristics probability fitting. The detailed content is as follows.
\subsection{Uncertainty Modeling and Batch Generation of Scenarios}
To more comprehensively characterize the randomness in the system's operation, it is necessary to consider additional factors, such as changes in scheduling methods and uncertainties in operating modes. GMM is a probability density function model composed of multiple normal distributions, and its mathematical expression is given in \eqref{eq10}~\eqref{eq12}.

\begin{equation}f_X(x)=\sum_{m=1}^M\omega_mN_m(x;\mu_m;\sigma_m)\label{eq10}\end{equation}
\begin{equation}\sum_{m=1}^M\omega_m=1,\omega_m>0\label{eq11}\end{equation}
\begin{equation}N_m(x;\mu_m;\sigma_m)=\dfrac{e^{-\dfrac{1}{2}(x-\mu_m)^T\sigma_m^{-1}(x-\mu_m)}}{(2\pi)^{D/2}det(\sigma_m)^{1/2}}\label{eq12}\end{equation}
Where:$M$ represents the total number of normal distributions; $\omega_m$, $\mu_m$, and $\sigma_m$ denote weights, means, and covariances of the $m$-th normal distribution. $N$ represents the probability density function of a normal distribution, $x$ is the independent variable (such as renewable power generation and other random variables), and $D$ is the dimension of independent variables.

Therefore, the first step is to use GMM to characterize uncertainty in power systems, generate parameter combinations via sampling, and subsequently obtain a set of power flow scenarios via batch simulation. Using GMM alone to model uncertainty can accurately capture the probability distribution of load and generation in real-world scenarios. However, it provides insufficient coverage for low-probability boundary cases, leading to poor generalization. Therefore, as a supplement, we use orthogonal tables to generate a complete set of parameter combinations for uncertain factors, thereby covering the sample distribution over the entire parameter space and enhancing the comprehensiveness of typical scenarios.

\subsection{Transient Data Generation and Scenarios Classification}
The second step describes the stability characteristics of the power flow scenarios using the transient stability index and classifies them by the stability characteristics of different scenario types. The specific idea is to introduce faults into power flow scenarios, run electromagnetic transient (EMT) simulations to obtain transient data, and subsequently compute stability indices. Subsequently, using stability indices as clustering features, the K-means method is applied to cluster power flow scenarios, and the resulting clusters are identified as those with similar stability characteristics.

Firstly, typical faults are set in power flow scenarios, or by traversing N-1 faults, stability of power flow scenarios under faults is observed through simulation, including rotor angle stability, voltage stability, and frequency stability. For this purpose, electrical quantities to be measured include bus voltage, generator rotor angle, and frequency. The calculation method of stability indices is as follows:

1)The Guidelines for Voltage Stability Evaluation of Power Systems \cite{a29} stipulate: "During the transient process after a large disturbance in power systems, the load bus voltage should recover to 0.80 p.u. within 10 seconds." Therefore, in this paper, the integral area of the voltage response curve below 0.80 p.u. is used to quantify the severity of voltage transient instability.

2) By using the bilinear inverse transformation method, the arc with radius $\pi$ on the complex plane in the rotating coordinate system is mapped to the interval [-1,1], as shown in \eqref{eq13}. When the generator rotor angle is spread out, TSI tends to -1.
\begin{equation}\text{TSI}_i=\dfrac{\pi-\delta_i}{\pi+\delta_i}\label{eq13}\end{equation}

3) The Rate of Change of Frequency (RoCoF) can quantitatively reflect the magnitude of the system's inertia and unbalanced power, and it responds faster than nadir.

Based on the stability discrimination indicators corresponding to the three types of stability above, these three indices can be used as clustering features for K-means clustering, thereby dividing the entire set of power flow scenarios into several clusters. The number of clusters should be determined based on the stability response characteristics.

\subsection{Typical Scenarios Generation based on System-level Characteristics Probability Fitting}
Based on the two steps above, power flow scenario clusters with distinct stability characteristics can be identified. For these power flow scenarios, Section \ref{222} defines and computes a set of system-level characteristics. For each obtained cluster of power flow scenarios, a system-level characteristics matrix is constructed. GMM is used to perform joint probability fitting on the system-level characteristics matrix of each cluster of power flow scenarios to obtain the joint probability density distribution. The power flow scenario closest to the peak of the probability density is selected as the typical scenario of the corresponding cluster.

\subsection{Application of Typical Scenarios}
The different scenario clusters obtained in the above steps have the corresponding stability characteristics. Therefore, stability can be predicted by the degree of closeness between the random scenario and the distribution of system-level characteristics across scenario clusters.

The joint probability density function of the system-level characteristics across different scenario clusters is fitted using GMM. The degree of attribution of any random scenario to each Gaussian component in the GMM is determined by the posterior probability, which depends on the Mahalanobis distance between the sample and the centers of each component. The Mahalanobis distance accounts for the covariance structure among features, as shown in \eqref{eq14}.

\begin{equation}D_m(x,\mu_m)=\sqrt{(x-\mu_m)^T\sigma_m^{-1}(x-\mu_m)}\label{eq14}\end{equation}
Where: $D_m$ represents the Mahalanobis distance of the system-level characteristics vector $x$ of the scenario relative to the $m$-th component in GMM.

In applications, to comprehensively consider the influence of multiple Gaussian components in GMM, weighted Mahalanobis distances can be introduced based on component weights:

\begin{equation}D_{\text{weighted}}=\sum_{m=1}^M\omega_mD_m(x,\mu_m)\label{eq15}\end{equation}

For the system-level characteristics vector $x$ corresponding to any power flow scenario, calculate its weighted Mahalanobis distance relative to each typical scenario cluster. The scenario cluster with the smallest distance is the one to which it belongs. Based on the stability of the scenario cluster, the scenario's stability under typical faults can be predicted.

\section{case study}\label{444}
\begin{figure}[tbp]
\centerline{\includegraphics[width=\columnwidth]{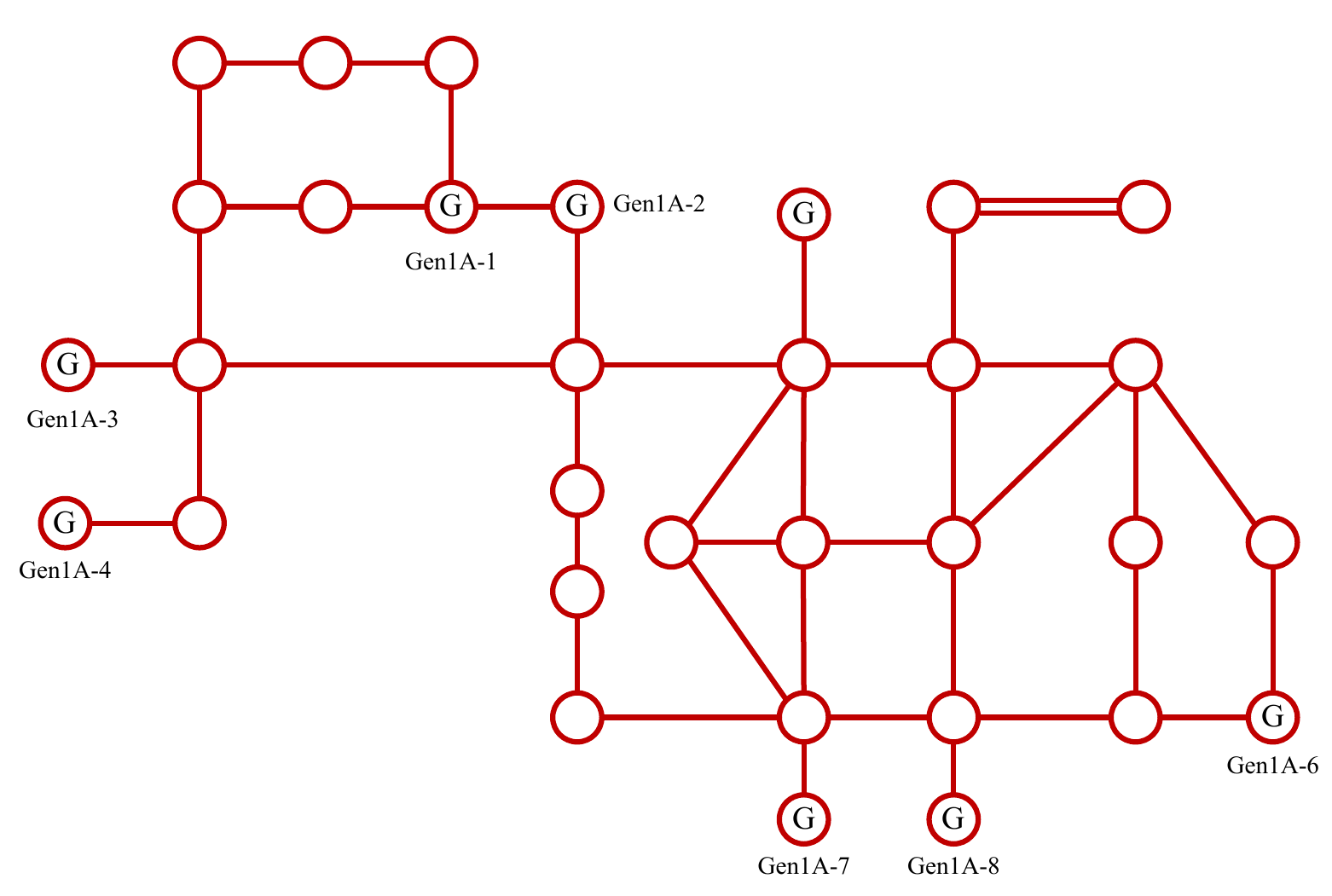}}
\caption{Topology of the CSEE benchmark}
\label{fig5}
\end{figure}

The test was conducted using the transient rotor angle instability coupling voltage collapse case released by CSEE. The calculation example comprises 97 nodes (45 nodes in the 500 kV main grid), and the topological connection diagram is shown in Figure \ref{fig5}. Installed capacity ratio: Renewable energy(2.4 GW): Synchronous generator(6.3 GW) =1:2.62. Renewable energy generation: 1.2 GW (wind power 0.6; photovoltaic 0.6), Synchronous generation: 2.49 GW. The system includes one HVDC link delivering 0.8 GW to the external grid. Among them, the modifiable uncertain factors include the penetration rate of renewable energy, power generation capacity, and power transmitted via HVDC.

\subsection{Electrical Coordinate System Construction and System-level Characteristics Calculation} 
\begin{figure}[tbp]
\centerline{\includegraphics[width=\columnwidth]{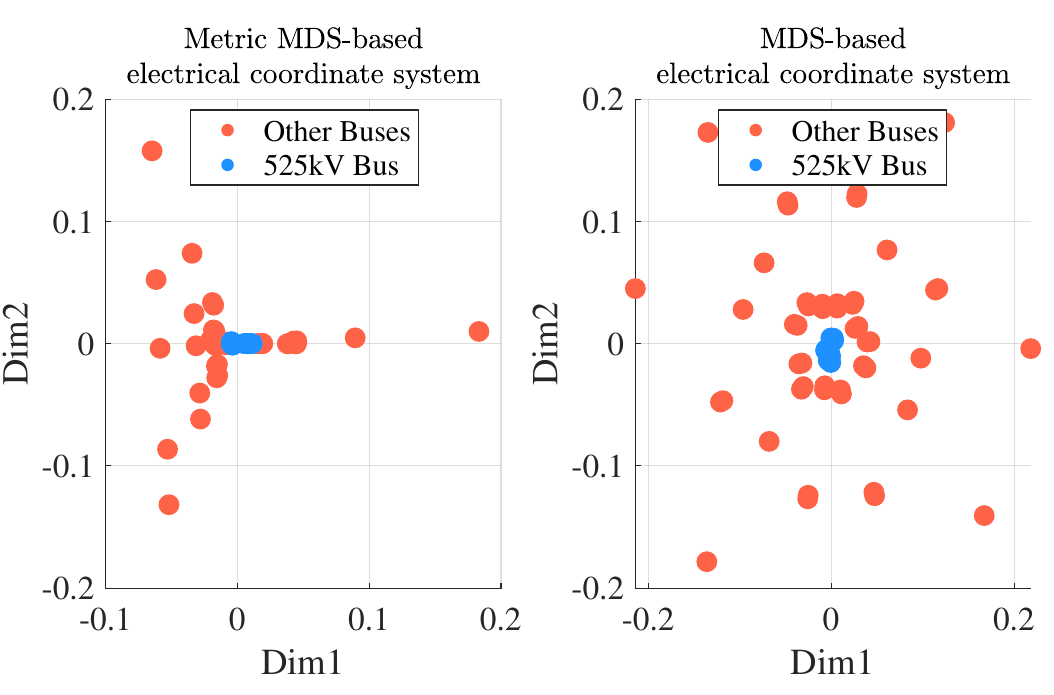}}
\caption{Electrical coordinate system of the CSEE}
\label{fig6}
\end{figure}

The electrical coordinate systems were constructed, respectively, based on classical MDS and metric MDS, and the first two-dimensional electrical coordinates were used, as shown in Figure \ref{fig6}. The electrical coordinate system maps the 500kV buses of the main grid at the center of the system, while lower-voltage buses are distributed on the periphery, which conforms to the actual topological structure of the power system. By comparing the two algorithms simultaneously, we find that, compared with the more accurate approximation of the numerical level of electrical distance provided by metric MDS, classical MDS more clearly depicts the nodes clustering characteristics of the power system.

\begin{figure}[tbp]
\centerline{\includegraphics[width=\columnwidth]{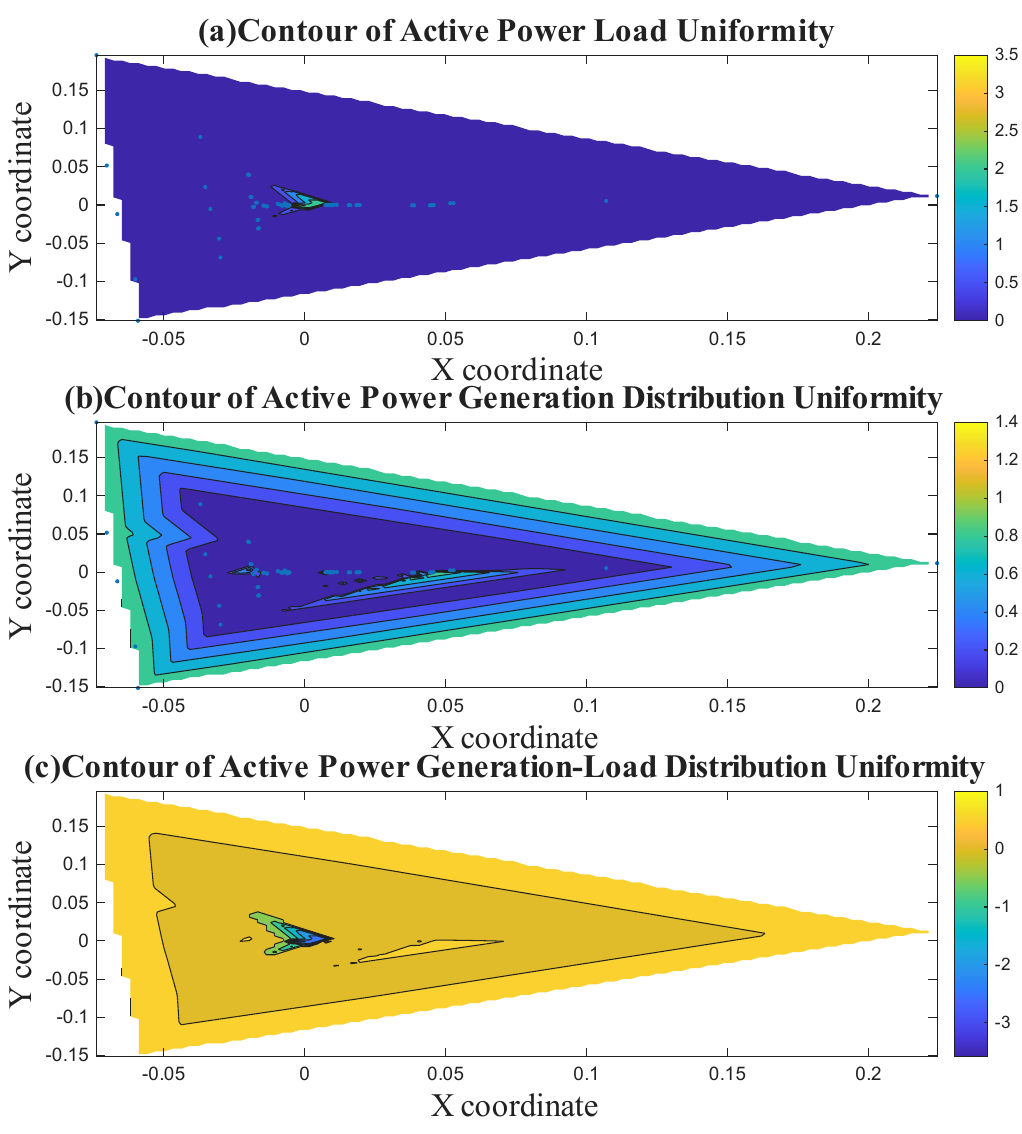}}
\caption{Active power generation–load distribution based on the electrical coordinate system}
\label{fig7}
\end{figure}

Taking the active power generation-load distribution as an example, the active power generation, load, and the difference between generation and load were introduced as new dimension values into the two-dimensional electrical coordinate system, and the generation-load distribution heat map is shown in Figure \ref{fig7}. According to the lower map of the active power generation-load distribution in Figure \ref{fig7}(c), it can be found that generation (the warm-colored part) is mainly distributed at the edge of the system, including renewable energy at the system boundary and synchronous generators at the center. Load (the cool-colored part) is concentrated at the center of the system, that is, the HVDC sending-end bus. Therefore, the generation-load distribution is highly unbalanced, and the power supply must be transported over long distances. When a branch fault prevents the successful transmission of power, a transient rotor angle instability coupling voltage collapse  phenomenon is likely to occur.

\subsection{Generation and Verification of Typical Scenarios}
\subsubsection{Generation of Scenarios with Uncertainties}

GMM is used to model uncertainty in four parameters: the penetration rate of renewable energy, the generation-load magnitudes of active/reactive power, and HVDC transmitted power, and to generate parameter combinations. At the same time, the orthogonal method is used to generate orthogonal combinations of parameters, enabling the generated samples to comprehensively cover the sample space and batch generate 471 trend scenarios.
\subsubsection{Scenarios Classification Based on Transient Stability Indices}

\begin{figure}[tbp]
\centerline{\includegraphics[width=0.7\columnwidth]{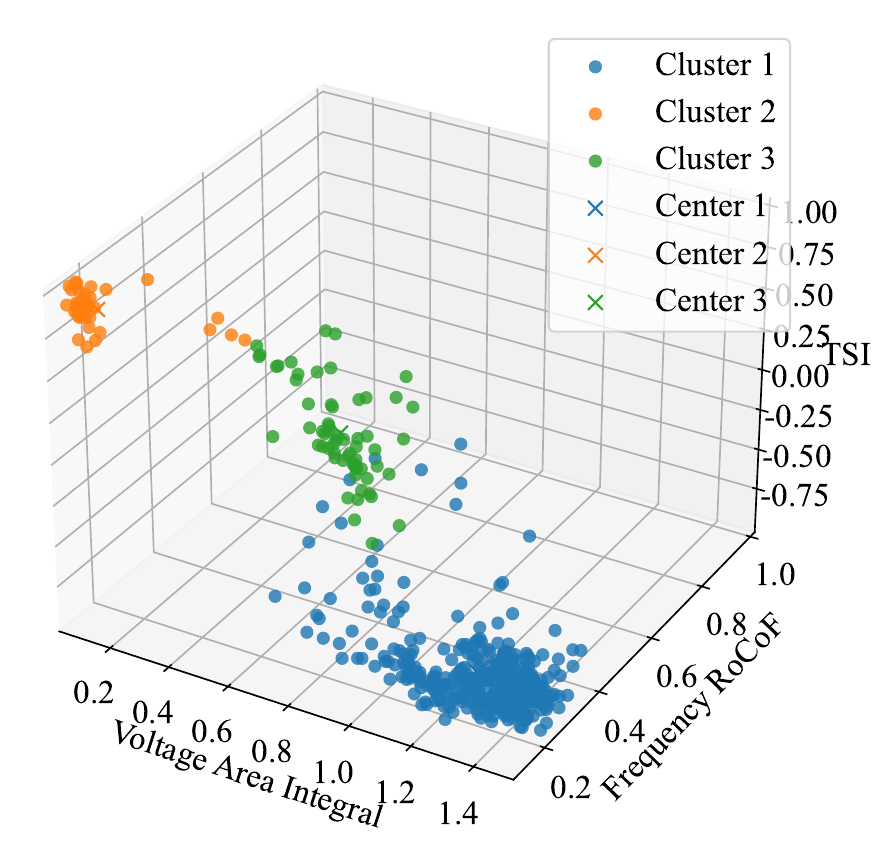}}
\caption{Clustered transient response characteristics}
\label{fig8}
\end{figure}

In the CSEE benchmark system, the typical fault is set to the three permanent N-1 faults on the transmission line between buses 1A-13 and 1A-12. Typical fault was set in power flow scenarios; EMT simulations were conducted, and the stability indices were obtained. The K-means clustering method was used to cluster the stability characteristics of the power flow scenarios, as shown in Figure \ref{fig8}. The overall trend exhibits three stability characteristics: stable, transient voltage instability, and transient rotor angle coupling-voltage instability.
\subsubsection{Typical Scenario Generation based on Distribution of System-level Characteristics}

For three different scenario clusters with distinct stability characteristics, calculate the system-level characteristics of scenarios within each cluster, respectively. GMM was used to fit the joint probability distribution of each cluster of macroscopic features, and ultimately, three typical scenarios were obtained. The specific operating parameters are shown in Table \ref{tab2}.

\begin{table*}
\centering
\caption{Data of typical scenarios}
\setlength{\tabcolsep}{3pt}
\begin{tabular}{>{\centering\arraybackslash}p{70pt}c>{\centering\arraybackslash}p{60pt}>{\centering\arraybackslash}p{50pt}>{\centering\arraybackslash}p{50pt}>{\centering\arraybackslash}p{50pt}>{\centering\arraybackslash}p{50pt}>{\centering\arraybackslash}p{50pt}}
\hline
Typical scenario &
MDS &
Renewable power generation(MW) & Active load(MW) & Reactive load(MW) & Reactive generation(MW) & HVDC delivered(MW) & Coverage rate\\
\hline
\multirow{2}{*}{stable scenario} & classic MDS & 639&	3531	&4442	&4498	&2924	&83.7\%\\ 
& metric MDS & 559	&2670	&4179	&3787	&1707&	79.1\% \\ \addlinespace
\multirow{2}{80pt}{transient rotor angle coupling voltage instability} & classic MDS & 1322	&7109	&4741	&5281	&2718	&72.6\%\\ 
& metric MDS & 1263	&7918	&4904	&5719	&3038	&56.5\% \\ \addlinespace \addlinespace
\multirow{2}{80pt}{transient voltage instability} & classic MDS & 777	&4813	&4462	&4548	&2082	&42.1\%\\ 
& metric MDS & 576	&5265	&4303	&4123	&2328	&30.9\% 
 \\ \hline
\end{tabular}
\label{tab2}
\end{table*}

\begin{figure}[tbp]
\centerline{\includegraphics[width=\columnwidth]{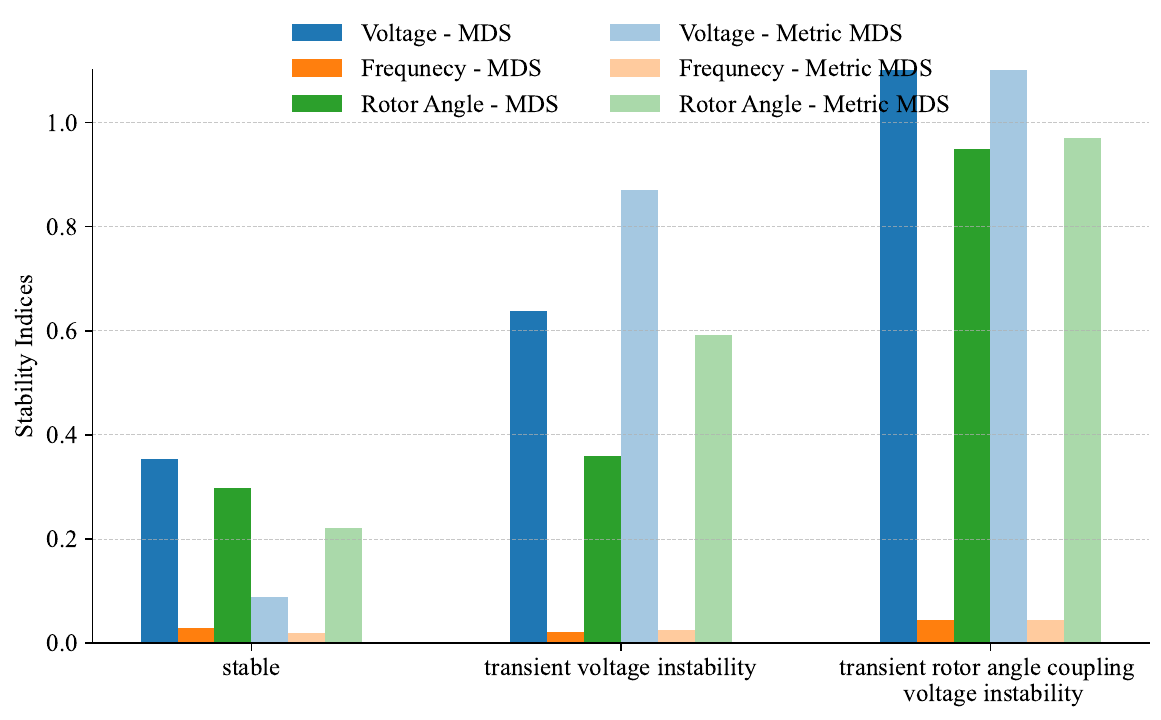}}
\caption{Verification results of fault traversal for typical scenarios}
\label{fig9}
\end{figure}

Verify the typicality of typical scenarios by two methods:

\textbf{a. Typicality of stability characteristics }: Results of stability indices for traversing N-1 faults in the obtained typical scenarios are shown in Figure \ref{fig9}. The larger the stability indices in the figure, the more severe the instability of the corresponding type of stability on average. Dark and light colors denote classic MDS and metric MDS, respectively. It can be observed that the stable typical scenario exhibits better stability across all three types of stability, whereas the transient voltage instability typical scenario and the transient rotor angle coupling voltage instability typical scenario are more prone to instability in their corresponding types of stability, indicating that the generated typical scenarios are typical of stability characteristics.

\begin{figure*}[tbp]
\centerline{\includegraphics[width=2\columnwidth]{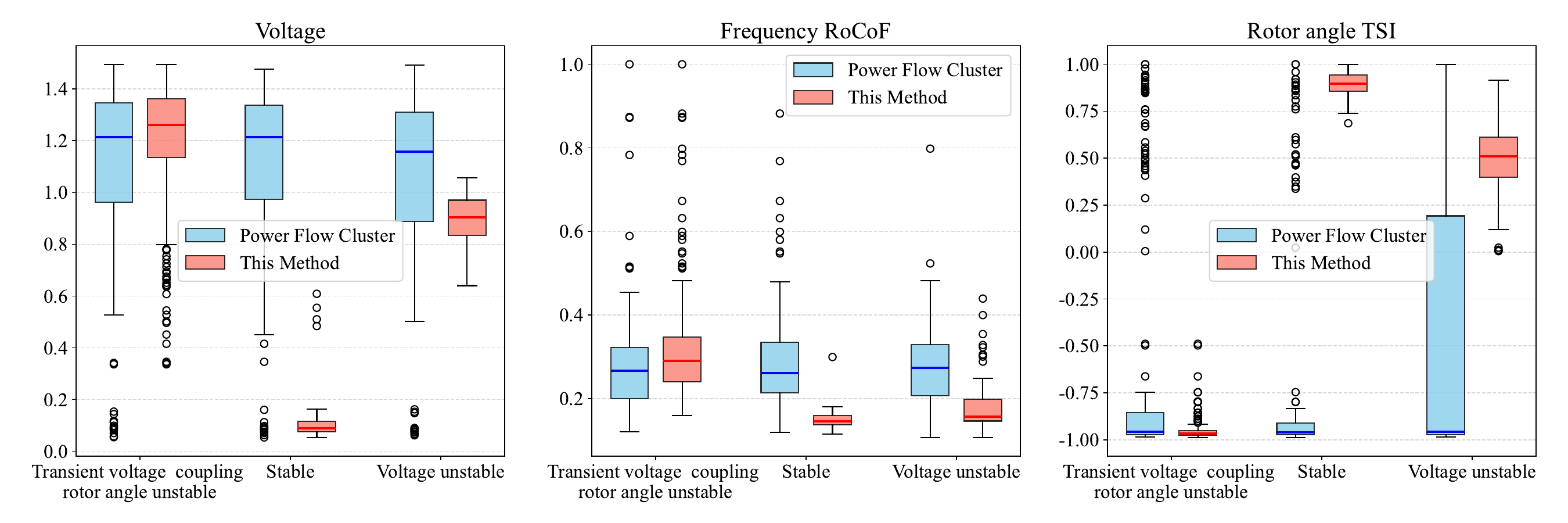}}
\caption{Stability indices distribution of scenario clustering}
\label{fig10}
\end{figure*}

The scenario classification method proposed in this paper, which considers the correlation between system-level characteristics and system stability, is compared with a traditional scenario classification method based on power flow and principal component analysis (PCA) clustering. The stability distribution for the scenario clusters is shown in Figure \ref{fig10}. The figure compares the classification results of two scenario classification methods for different stable types. It can be observed that the scenario clusters obtained by traditional PCA clustering based on power flow data cannot distinguish among different clusters. Compared with the classification results of the stability indices clustering, there is no overlap in stability indices across the three scenario clusters.

\textbf{b. Representativeness}: Use the coverage rate to quantitatively assess the representativeness of typical scenarios; if the distance between a given scenario and a typical scenario is less than the given threshold $\delta = 0.2$, it is considered covered. The proportion of scenarios covered by each typical scenario among all scenarios is defined in \eqref{eq16}.

\begin{equation}R_{cov}=\dfrac{N_{\text{cover}}}{N}\label{eq16}\end{equation}
Where: $N_{cover}$ is the number of covered scenarios, $N$ is the total number of scenarios.

The typical scenario coverage rates for the two algorithms, classic MDS and metric MDS, are reported in Table \ref{tab2}. We find that classic MDS is more representative of typical scenarios than metric MDS, indicating that a more accurate characterization of the node clustering is conducive to generating typical scenarios.

\subsection{Scenario Stability Prediction based on Typical Scenarios}
According to the clustering results in Figure \ref{fig8}, all test scenarios can first be divided into two categories: unstable and stable. The unstable scenarios can be further classified into two categories based on their stability types. Therefore, the prediction of stability can be divided into two stages, namely:

1) Determine whether the scenario will become unstable under disturbance.

2) Determine whether the unstable scenario will experience severe instability due to the coupling of transient rotor angle instability and voltage collapse.

A total of 65 test scenarios were randomly generated using GMM, and the corresponding system-level characteristics matrices were calculated. The weighted Mahalanobis distance is used to discriminate the subordinate attributes of three scenario clusters in a random scenario. The precision and recall rates for the two discrimination links are shown in Table \ref{tab3}. The precision and recall rates of the classic MDS algorithm are both above 90\%, significantly higher than those of the metric MDS algorithm. It fully demonstrates the effectiveness of the typical scenario generation methods and the superiority of the classical MDS algorithm over metric MDS in capturing the system-level characteristics of the power system. At the same time, it can achieve 100\% identification of scenarios with potential instability risks, meeting the engineering requirements for power grid safety.
\begin{table}
\caption{Data of typical scenarios}
\setlength{\tabcolsep}{3pt}
\begin{tabular}{>{\centering\arraybackslash}p{40pt}>{\centering\arraybackslash}p{80pt}>{\centering\arraybackslash}p{50pt}>{\centering\arraybackslash}p{50pt}}
\hline
MDS &
Prediction stage & Precision & Recall\\
\hline
\multirow{2}{*}{classic MDS} & stability & 92.73\%&	100\%\\ 
& whether the instability is coupled& 97.37\%	&90.24\% \\ \addlinespace
\multirow{2}{*}{mertic MDS} & stability & 85.00\%&	100\%\\ 
& whether the instability is coupled& 83.72\%	&87.80\%
 \\ \hline
\end{tabular}
\label{tab3}
\end{table}
\section{Conclusion}
This paper addresses the limited guidance provided by typical scenarios for stability analysis in power-electronics-dominated power systems and proposes a method for generating such scenarios and predicting system stability based on the system's system-level characteristics. The specific conclusions are as follows:

1) The electrical coordinate system based on the classic MDS can precisely depict the topological structure of the system and the clustering characteristics of nodes. Meanwhile, as the dimension increases, the electrical distance can be approximated numerically.

2) The system-level characteristics defined based on the electrical coordinate system have a strong correlation with both the power flow characteristics and the stability characteristics of the system. The use of kernel functions can more accurately reflect the nonlinear correlations among them.

3) The typical scenarios based on the classic MDS are typical in both the typicality and representativeness of stability characteristics. The method based on the weighted Mahalanobis distance can predict scenario stability and achieve a 100\% recognition rate for unstable scenes.

Currently, the definition of system-level characteristics relies on power flow data. Using steady-state data obtained from power flow calculations to predict transient data under fault conditions has significant theoretical limitations. Therefore, it is necessary to further refine the theoretical derivation and integrate it with the control characteristics of power electronic equipment, such as renewable energy sources.

\textbf{Tao Li} received the B.E. degree in electrical engineering from
Tsinghua University, Beijing, China, in 2024. She
is currently working toward the Ph.D. degree in electrical engineering at Tsinghua University, Beijing,
China. Her research interests include stability analysis of power  systems with renewable energy.

\textbf{Chen Shen}(Senior Member, IEEE) received the B.E.
and Ph.D. degrees in electrical engineering from Tsinghua University, Beijing, China, in 1993 and 1998,
respectively. From 1998 to 2001, he was a Postdoc
Research Fellow with the Department of Electrical
Engineering and Computer Science, University of
Missouri Rolla, Rolla, MO, USA. From 2001 to 2002,
he was a Senior Application Developer with ISO New
England, Inc., Holyoke, MA, USA. Since 2009, he
has been a Professor with the Department of Electrical
Engineering, Tsinghua University. He is currently the
Director of the Energy Digitization Research Center, Sichuan Energy Internet
Research Institute, Tsinghua University. He has authored or coauthored more
than 180 technical papers and one book, and holds 32 issued patents. His
research interests include power system analysis and control, renewable energy
generation, and smart grids.

\end{document}